\documentclass[twocolumns]{aa} 

\usepackage{graphicx}
\usepackage{float}
\usepackage{subfig}
\usepackage{txfonts}
\usepackage{comment}
\usepackage{url}
\usepackage[switch]{lineno}
\usepackage{datetime}
\usepackage[usenames,table,dvipsnames]{xcolor}
\usepackage{setspace}
\usepackage[normalem]{ulem}
\usepackage{soul}
\definecolor{blue}{RGB}{66, 153, 233}
\definecolor{red}{RGB}{255, 0, 0}
\definecolor{purple}{RGB}{255, 0, 255}

\newcount\Comments  
\newcommand{\kibitz}[2]{\ifnum\Comments=1\textcolor{#1}{#2}\fi}

\begin{document}

   \title{Comparing the three-dimensional morphological asymmetries in the ejecta of Kepler and Tycho in X-rays}
   \titlerunning{Morphological asymmetries in the ejecta of Cassiopeia A in X-rays}
   \authorrunning{Please Picquenot et al.}
   \author{A. Picquenot$^{1,2,3}$
          \and
          T. Holland-Ashford$^{4}$
          \and
          B.~J. Williams
          \inst{2}
          }
 

   \institute{Department of Astronomy, University of Maryland, College Park, MD 20742 
              \and
              X-ray Astrophysics Laboratory NASA/GSFC, Greenbelt, MD 20771 
              \and
              Center for Research and Exploration in Space Science and Technology, NASA/GSFC, Greenbelt, MD 20771, USA
              \and
              Center for Astrophysics | Harvard \& Smithsonian, 60 Garden St, Cambridge MA 02138, USA
   }
   

   \date{\today}

 
  \abstract
  {}
  {Recent simulations have shown that asymmetries in the ejecta distribution of supernova remnants (SNRs) may be a reflection of asymmetries left over from the initial supernova explosion. Thus, SNR studies provide a vital means for testing and constraining model predictions in relation to the distribution of heavy elements, which are key to improving our understanding of the explosion mechanisms in Type Ia supernovae.}
  {The use of a novel blind source separation method applied to the megasecond X-ray observations of the historic Kepler and Tycho supernova remnants has revealed maps of the ejecta distribution. These maps are endowed with an unprecedented level of detail and clear separations from the continuum emission. Our method also provides a three-dimensional (3D) view of the ejecta by individually disentangling red- and blueshifted spectral components associated with images of the Si, S, Ar, Ca, and Fe emission. This approach provides insights into the morphology of the ejecta distribution in those two remnants.}
  {Those mappings have allowed us to thoroughly investigate the asymmetries in the intermediate-mass elements and Fe distribution in two Type Ia supernova remnants. We also compared the results with the core-collapse Cassiopeia A remnant, which we had studied previously. The images obtained confirm, as expected for Type Ia SNRs, that the Fe distribution is mostly closer to the core than that of intermediate-mass elements. They also highlight peculiar features in the ejecta distribution, such as the Fe-rich southeastern knot in Tycho.}
  {}
   


   \maketitle
%

\section{Introduction}

The supernovae (SNe) of 1572 (hereafter, Tycho) and 1604 (hereafter, Kepler) are undoubtedly two of the most famous historical supernova remnants (SNRs), associated with  "stella novae" events, as documented in the literature \citep{Tycho73,kepler06}. Both have left a lasting astronomical and philosophical impact. However, despite their historical importance, the ignition mechanisms and the deflagrating or detonating nature of the nuclear burning front in Type Ia supernovae (SNe), including both Tycho and Kepler, are still poorly understood. Discriminating among the single-degenerate (a white dwarf accretes matter from a companion star) and the double-generate (explosion caused by the merging of two white dwarfs) scenarios is still a challenge and we do not have a precise idea of the relative quantity of single- and double-degenerate explosions.

In this context, three-dimensional (3D) simulations have shown that asymmetries are a key part of the explosion mechanisms \citep[e.g.,][]{Seitenzahl2013,Stone_2021}. It has been argued that the diversity seen in the spectral evolution of Type Ia supernovae may be a consequence of the viewing angle on an asymmetric explosion \citep{Maeda10}. Recently, simulations presenting the evolution of a Type Ia SNR over a period spanning one year after the explosion to several centuries afterward have shown that asymmetries present in the original SN could still be observed for an extended amount of time \citep{Ferrand_2019,Ferrand_2022}.

As a consequence, the study of the ejecta asymmetries in remnants from Type Ia SNe is a logical way to probe simulations predictions and constrain explosion mechanisms. In particular, models predict that Fe-group elements (e.g., Cr, Mn, Ti, Fe, Ni) are located in the inner parts of the ejecta, surrounded by an outer shell of intermediate-mass elements (e.g. Si, S, Ca and Ar); the same models predict the velocity of intermediate-mass elements to be higher on average than that of the Fe-group \citep{Seitenzahl2013,Stone_2021}. While \cite{Miceli_2015} confirmed this expected spatial distribution in Tycho using {\it XMM-Newton} observations, \cite{Stone_2021}  discussed the case of G1.9+0.3, which presents surprisingly high-velocity Fe emission. 

As young, bright objects benefiting from  deep 
 {\it Chandra} observations, Tycho and Kepler, both  are prime targets for studying ejecta asymmetries in type Ia SNRs. Tycho has been classified as a type Ia based on its observed light curve and color evolution \citep{Ruiz04}, and it is estimated to be at a distance of $\sim 2.8$ kpc \citep{Kozlova_2018}. It is thought to originate from a single-degenerate scenario, which is supported by the discovery of a star near the center of the remnant \citep{Ruiz04}. Its environment is known to be inhomogeneous and Tycho shows traces of interactions with the circumstellar medium \citep[CSM;][]{Lu2011,Williams_2013}. \cite{Zhou_2016} reports the observation of a clumpy molecular bubble surrounding Tycho, which also supports a single-degenerate scenario to explain its origin.

Kepler's type Ia identification is mainly based on X-ray observations showing shocked ejecta with strong Si, S, and Fe emission and a near absence of O emission \citep{Kinugasa99,Cassam04,Reynolds07}. It is estimated to be at a distance of $\sim 5$ kpc \citep{Sankrit16} and its relatively young age indicates that most of the emission is due to heated-up ejecta material rather than swept-up interstellar medium \citep[ISM ;][]{Kinugasa99,Cassam04}. However, it seems to be interacting with the dense CSM \citep[]{Blair_2007,Williams_2012}, which would favour a single-degenerate scenario \citep{Patnaude_2012}. Nevertheless, a companion star is yet to be found \citep{Kerzendorf_2014,Sato_2017,Ruiz-Lapuente_2018}

The aim of this paper is to present a thorough study of the morphological asymmetries in the Si, S, Ca, Ar, and Fe in Kepler and Tycho's ejecta. Using a blind source separation method (BSS), we were able to extract detailed and unpolluted maps associated with a mean spectrum for each individual elements, which we used to quantify and compare the morphological and velocity asymmetries. In Section \ref{sect:method}, we describe our method. In Sections \ref{sect:Kepler} and \ref{sect:Tycho}, we present and discuss our results for Kepler and Tycho, respectively.  Finally, since we are applying the same methodology as we did in \cite{Picquenot_2021} for the Cassiopeia A core-collapse supernova, we compare the asymmetries we observed in the Tycho, Kepler, and Cassiopeia A SNRs  in Section \ref{sub:comparison}.


\begin{table}
    \centering

    \renewcommand{\arraystretch}{1.4}
    \begin{tabular}{c c c c c}

   Object & ObsID & Date & ACIS-S & ACIS-I \\ 
    \hline 
 Kepler & 116 & 2000 Jun 30 \ & 48.8 & -
 \\ 
  & 4650 & 2004 Oct 26 \ & 46.2 & -
 \\ 
   & 6714 & 2006 Apr 27 \ & 157.8 & -
 \\ 
  & 6715 & 2006 Aug 3 \ & 159.1 & -
 \\ 
  & 6716 & 2006 May 5 \ & 158.0 & -
 \\ 
  & 6717 & 2006 Jul 13 \ & 106.8 & -
 \\ 
  & 6718 & 2006 Jul 21 \ & 107.8 & -
 \\ 
  & 7366 & 2006 Jul 16 \ & 51.5 & -
 \\ 
 & 16004 & 2014 May 13 \ & 102.7 & -
 \\ 
 & 16614 & 2014 May 16 \ & 36.4 & -
 \\ 
 \cline{3-5}
 & & Total & 975.1 & - \\
\hline
Tycho & 115 & 2000 Sep 20 & 48.9 & - 
 \\ 
 & 3837 & 2003 Apr 29 & - & 145.6 
 \\ 
 & 7639 & 2007 Apr 23 & - & 108.9 
 \\ 
 & 8551 & 2007 Apr 26 & - & 33.3 
 \\ 
 & 10093 & 2009 Apr 13 & - & 118.3 
 \\ 
 & 10094 & 2009 Apr 18 & - & 90.0 
 \\
 & 10095 & 2009 Apr 23 & - & 173.4 
 \\
 & 10096 & 2009 Apr 27 & - & 105.7 
 \\
 & 10097 & 2009 Apr 11 & - & 107.4 
 \\
 & 10902 & 2009 Apr 15 & - & 39.5 
 \\
 & 10903 & 2009 Apr 17 & - & 23.9 
 \\
 & 10904 & 2009 Apr 13 & - & 34.7 
 \\
 & 10906 & 2009 May 3 & - & 41.1 
 \\
 & 15998 & 2015 Apr 22 & - & 147.0 
 \\
 & 23538 & 2021 Sep 27 & - & 28.7 
 \\
 & 23541 & 2021 Oct 16 & - & 28.1 
 \\
 & 24391 & 2021 May 29 & - & 28.4 
 \\
 \cline{3-5}
 & & Total & 48.9 & 1,254.0 \\
\hline 
\end{tabular}
\caption{\label{tab:data-description}Data from {\it Chandra} used in our study.}
\label{sect:obs}
\end{table}

\section{Method}
\label{sect:method}

To retrieve accurate maps of the main thermal line components in Kepler and Tycho, we used a BSS method based on the General Morphological Components Analysis \citep[GMCA, see][]{bobin15}, first introduced for X-ray observations by \cite{picquenot:hal-02160434}. This algorithm can disentangle spectrally and spatially mixed components from an X-ray data cube of the form $(x,y,E)$. In particular, the algorithm has proven  capable of extracting extremely faint components from X-ray data cubes. An updated version of this algorithm, the pGMCA \citep[see][]{9215040}, has been developed to take into account the Poissonian nature of X-ray data. It was first used on Cassiopeia~A data from {\it Chandra, } demonstrating its suitability for producing clear, detailed, and unpolluted images of both the ejecta \citep{Picquenot_2021} and the synchrotron emission at different energies \citep{picquenot23}. Given the pGMCA algorithm is primarily designed to focus on spatial morphology, the extracted spectra are sometimes imperfectly reconstructed; however, they can usually be used to assess the retrieved components' nature and physical properties.

\subsection{Image analysis}
\label{sect:image}

In this paper, we used pGMCA on stacked {\it Chandra} observations of Kepler and Tycho (see Table \ref{sect:obs}) with specific binnings to increase the photon counts in each bin. For Kepler,  we used a $16$" spatial binning and a $43.8$ eV energy binning. For Tycho, we used a $30$" spatial binning and a $43.8$ eV energy binning.

We applied the pGMCA algorithm on the resulting $(x,y,E)$ data cubes of Kepler and Tycho around five energy bands surrounding main emission lines: Si ($1.5 - 2.2$ keV), S ($2.0 - 2.7$ keV), Ar ($2.7 - 3.7$ keV), Ca ($3.4 - 4.4$ keV), and Fe ($5.0 - 8.0$ keV). We were able to retrieve maps of their spatial distribution associated with spectra, successfully disentangling them from the synchrotron emission or other unwanted components. The elements emission distribution were retrieved as different images associated with spectra that we interpret as being the same emission lines slightly red or blueshifted. We labelled the components with the most extreme shifts as  ``redshifted'' and ``blueshifted.'' We conveniently labeled the other components as ``other.'' We note that with pGMCA, the number of components to retrieve is user-defined. After several tests, we chose the number that gave the most morphologically different and physically meaningful components for each energy line.

We then applied the power-ratio method (PRM) to quantitatively analyze and compare the asymmetries in the images extracted by the pGMCA algorithm. The PRM was developed by \citet{1995ApJ...452..522B} and previously employed for use on SNRs \citep{lopez09b,Lopez_2009,lopez11,Picquenot_2021}. It consists of calculating multipole moments in a circular aperture positioned on the centroid of the image, with a radius that encloses the whole SNR. Powers of the multipole expansion $P_m$ are then obtained by integrating the $m$th term over the circle. To normalize the powers with respect to flux, they are divided by $P_0$, thus forming the power ratios $P_{m}/P_0$. For a more detailed description of the method, see \citet{Lopez_2009}.

The $P_2/P_0$ and $P_3/P_0$ terms convey complementary information about the asymmetries in an image. The first term is the quadrupole power-ratio and quantifies the ellipticity and elongation of an extended source, while the second term is the octupole power-ratio and is a measure of mirror asymmetry. Both are to be compared simultaneously to ascertain the asymmetries in different images.

To compare asymmetries in the blue- and redshifted parts of the elements' distribution, we proceeded as in \cite{Picquenot_2021}. In a first step, we calculated the $P_2/P_0$ and $P_3/P_0$ ratios of each element's total distribution by using the sum of the shifted components maps as an image, whose centroid is used as an approximation of the center-of-emission of the considered element. Then, we calculated the power ratios of the blue- and redshifted images separately using the same center-of-emission. Ultimately, we normalized the power ratios thus obtained by the power ratios of the total element's distribution:

\begin{equation}
P_i/P_{0\textit{  (shifted / total)}} =\frac{P_i/P_{0\textit{  (red or blue image)}}}{P_i/P_{0\textit{  (total image)}}},
\end{equation}
\label{eq:Pshifted}

\noindent
where $i=2$ or $3$ and $P_i/P_{0\mbox{ (red or blue image)}}$ is calculated using the centroid of the total image. That way, we can compare the relative asymmetries of the shifted parts of different elements, without the comparison becoming biased by the original asymmetries of the whole distribution.

Finally, to obtain an estimation of the errors in our analysis, we used the constrained bootstrap introduced in \cite{Picquenot_2021}. This method does not produce physically meaningful errors, but produces resamplings of the original data to assess the robustness of the pGMCA algorithm. Unfortunately, the algorithm is so sensitive to initial conditions that on some resamplings, it often fails to capture the same components found in the original data. For that reason, we defined our error bars as the standard deviation of the $50$ best results out of $110$ resamplings for each component. The resulting errors constitute a tentative attempt at estimating the uncertainties when using an advanced analysis tool, such as pGMCA, as a general and non-trivial problematic. We note that the error values tend to increase with the total number of counts, as more counts means more potential components to disentangle, which reduces the algorithm's robustness. 

\subsection{Spectral analysis}
\label{sect:image}

Although we classified the extracted pGMCA components by their 2D morphologies and spectra, the differences in line centroid for each component do not necessarily correspond to the plasma's Doppler shift. The ionization state of the plasma also affects spectral features: lower ionization plasmas typically have lower-energy emission lines. Thus, in order to robustly classify certain components as ``redshifted'' or ``blueshifted,'' we needed to disentangle the effects of ionization and Doppler shifts. To do so, we fit the plasmas using {\it Xspec} (version 12.13.0e). For each component, we simultaneously fit the background-subtracted emission from all elements using a complex plasma model \texttt{TBabs*(vpshock$_{Si-Ca}$+vvpshock$_{Fe}$+powerlaw)}, which 
includes a \texttt{vpshock} component with variable abundances for the Si-Ca emission, a \texttt{vvpshock} component to capture the Fe K-$\alpha$ and surrounding element emission, along with a power law to capture any remaining nonthermal emission in these components. The uncertainties for best-fit parameters reported reflect the 90\% confidence interval as determined by the Xspec \texttt{error} command. We used the redshift parameter as a proxy for Doppler shift. Essentially, we wanted to ensure that each of our identified components have distinct velocities, and that the shifts we observe cannot be explained as similar plasmas in different ionization states. This process also serves as a double check that the pGMCA method is extracting physically plausible emission and can be understood as shock-heated ejecta, ISM, or synchrotron emission. The spectra extracted by pGMCA are sometimes imperfectly reconstructed and the resulting fits can be burdened by large errors.

\begin{figure}
\centering
\includegraphics[width=9cm]
{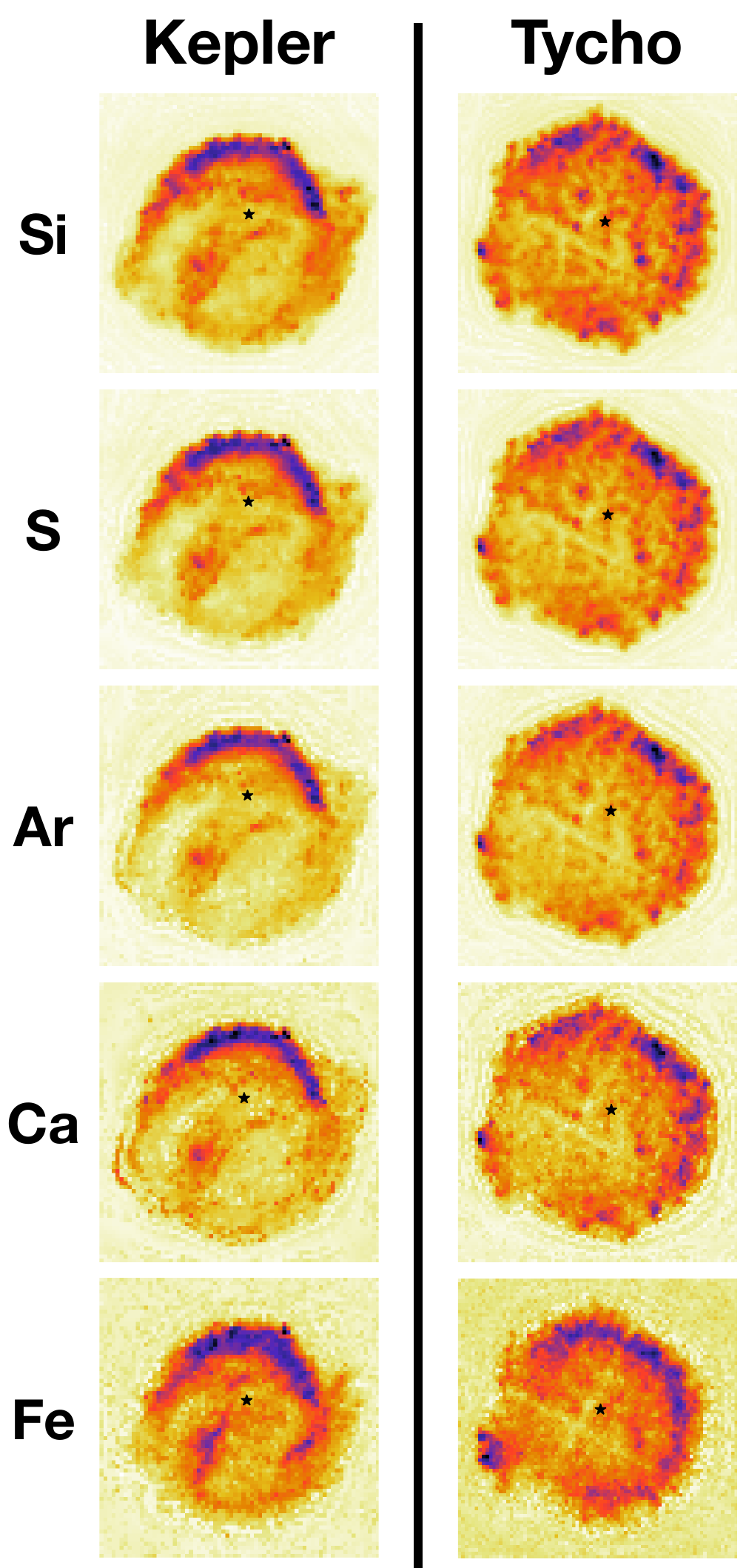}
\caption{Total images of the different line emission spatial structure as retrieved by pGMCA.  Left: For Kepler.  Right: For Tycho. The black star represents the image centroid adopted in the PRM analysis. The color scale is presented as a square root.}
\label{fig:red_blue_images-TOT}
\end{figure}
\section{Results}
\label{sect:results}

\begin{figure*}
\centering
\includegraphics[width=17.6cm]
{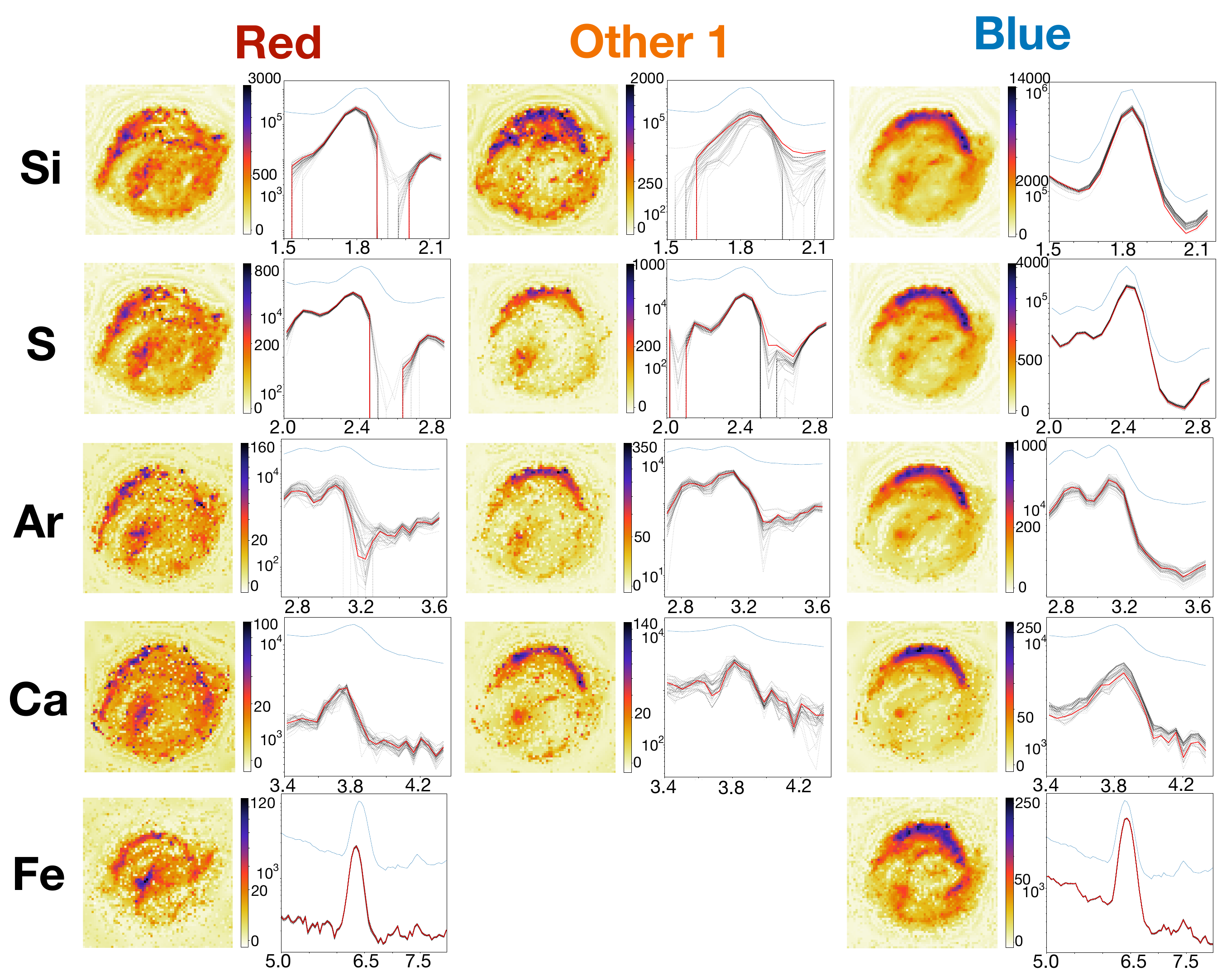}
\caption{Red, blue-shifted, and ``other'' parts in the Si, S, Ar, Ca, and Fe spatial distribution and their associated spectrum as found by pGMCA in Kepler. The spectra in red correspond to the application of the algorithm on real data, while the dotted gray spectra correspond to 50 selected applications on constrained bootstrap resamplings, illustrating statistical uncertainties. The x-axis is in keV and the y-axis in counts.}
\label{fig:red_blue_images-Kepler}
\end{figure*}

\begin{figure*}
\centering
\includegraphics[width=17.6cm]
{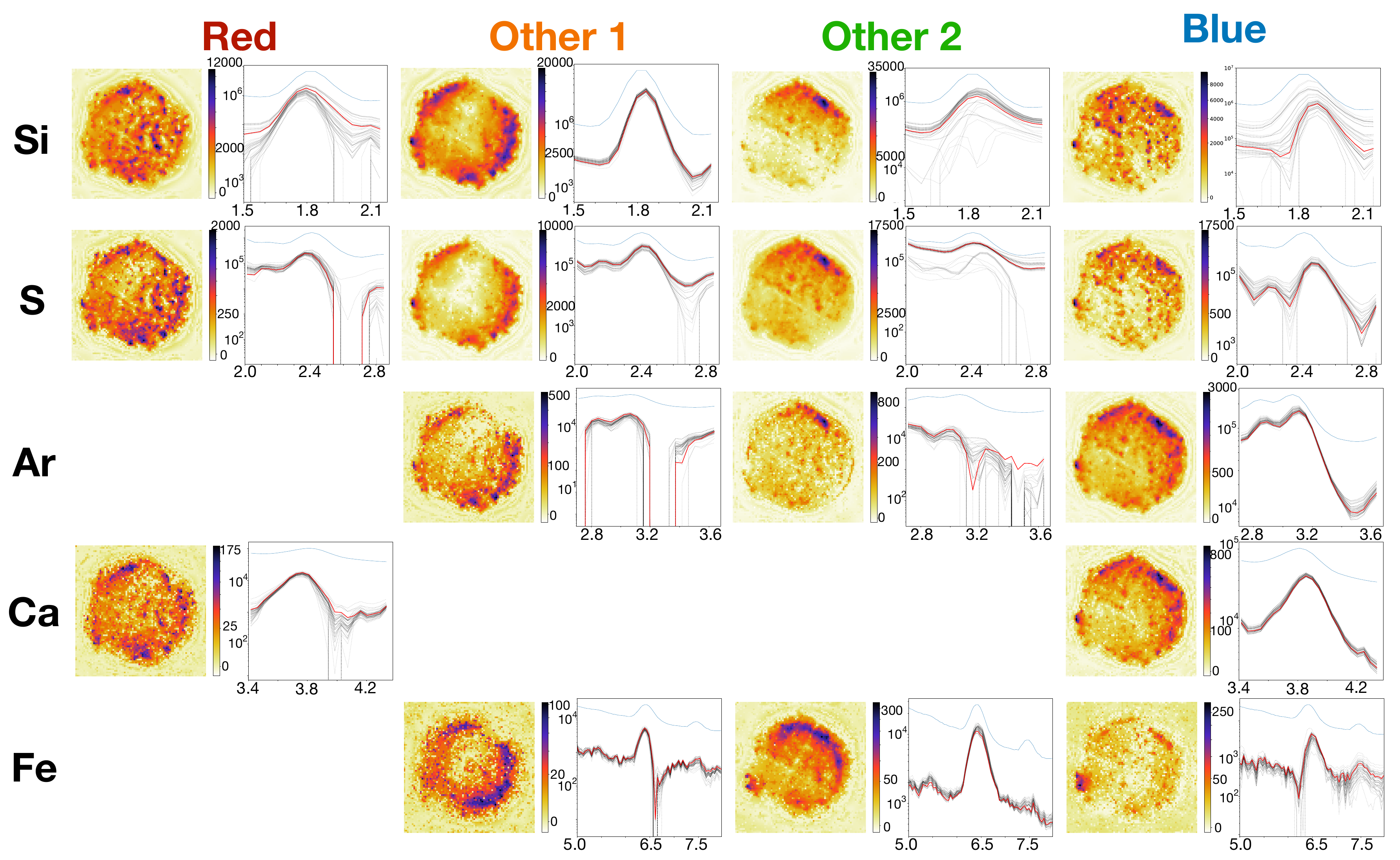}
\caption{Red-, blueshifted, and ``other'' parts in the Si, S, Ar, Ca, and Fe spatial distribution and their associated spectrum as found by pGMCA in Tycho. The spectra in red correspond to the application of the algorithm on real data, while the dotted gray spectra correspond to fifty selected applications on constrained bootstrap resamplings, illustrating statistical uncertainties. The x-axis is in keV and the y-axis in counts.}
\label{fig:red_blue_images-Tycho}
\end{figure*}

Figure~\ref{fig:red_blue_images-TOT} shows the total images for all five line emission in both remnants, obtained by summing the shifted parts. It also indicates the centroid of each image that is adopted in the PRM. In Kepler, Fig.~\ref{fig:red_blue_images-Kepler} shows the differently shifted parts together with their associated spectra, Fig.~\ref{fig:red_blue-Kepler} shows the PRM plot, and Fig.~\ref{fig:Kepler-xspec} gives the results of the spectral analysis. In Tycho, Fig.~\ref{fig:red_blue_images-Tycho} shows the differently shifted parts, Fig.~\ref{fig:red_blue-Tycho} shows the PRM plot, and Fig.~\ref{fig:Tycho-xspec} givds the results of the spectral analysis. From the images of Fig.~\ref{fig:red_blue_images-Kepler} and Fig.~\ref{fig:red_blue_images-Tycho}, we  derived the ratios presented in Table \ref{tab:frac-Kepler} (for Kepler) and Table \ref{tab:frac-Tycho} (for Tycho), which highlight the relative importance of each component in the total line emission. Table \ref{tab:fits-xspec} presents the best fit parameters from the spectral analysis from both remnants.

\begin{table}
\centering
\begin{tabular}{c|c c c}
  \hline
  \hline
    & Redshifted & Other 1 & Blueshifted \\
  \hline
  Si & 0.20 & 0.14 &  0.66 \\
  
  S & 0.19 & 0.08 & 0.73 \\
  
  Ar & 0.14 & 0.17 & 0.69 \\
  
  Ca & 0.30 & 0.23 & 0.47 \\
  
  Fe & 0.26 & - & 0.74 \\
  \hline
\end{tabular}
\caption{Fractions of the counts in the total image that belong to the redshifted, the blueshifted or the ``other'' parts, for each line in Kepler.}
\label{tab:frac-Kepler}
\end{table}

\subsection{Kepler}
\label{sect:Kepler}

In Kepler, the line emission total images from Fig.~\ref{fig:red_blue_images-TOT} appear very similar to each other, with the exception of Fe that shows a more compact distribution, generally closer to the geometrical center. The shifted images from Fig.~\ref{fig:red_blue_images-Kepler} also show a great similitude between the differently shifted components for each line emission. The most blue-shifted component always appears as a broad rim following the forward shock in the north, and the ``other'' component follows a similar but thinner distribution, closer to the edge near the shock and with stronger emission in the west. The most extreme red-shifted component is more diffuse, with peaks present in the north-east, the west and slightly to the east of the geometrical center. The red component from the Fe line is significantly different, with a clear peak near the center and comparatively less prominent features other than the northern rim. The Si ``other'' component is more complex and shows more features than the northern rim dominating ``other'' components in S, Ar, and Ca emission. This might be due to the higher number of counts at lower energies, which would allow pGMCA to reconstruct an image endowed with more details.

\begin{figure}
\centering
\includegraphics[width=9cm]
{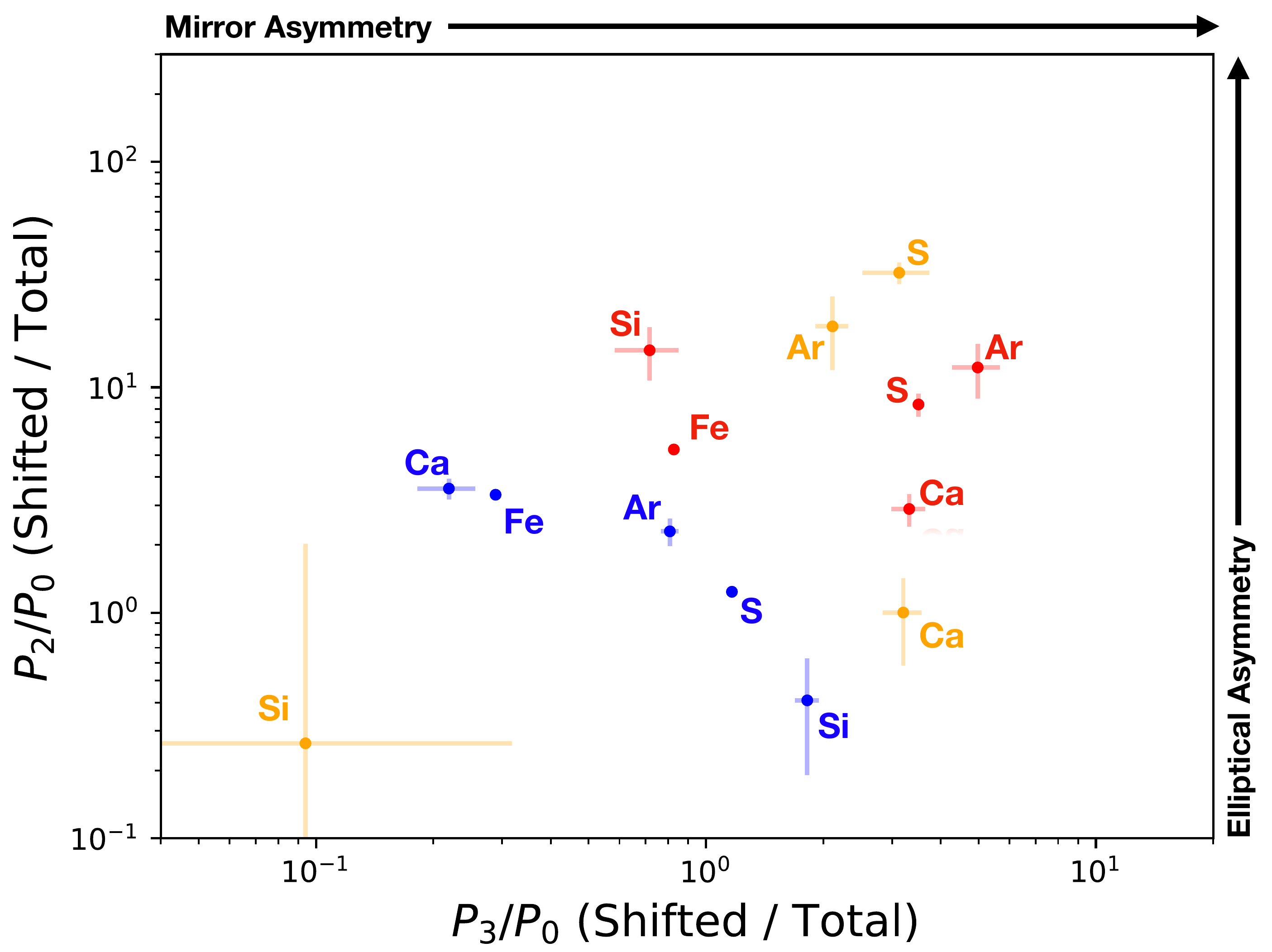}
\caption{Quadrupole power-ratios $P_{2}/P_{0}$ versus the octupole power-ratios $P_{3}/P_{0}$ of the red, blueshifted, and ``other'' images of the different line emission in Kepler shown in Fig. \ref{fig:red_blue_images-Kepler}, normalized with the quadrupole and octupole power-ratios of the total images. The dots represent the values measured for the pGMCA images obtained from the real data, and the error bars the standard deviation of fifty selected outputs from pGMCA on constrained bootstrap resamplings.}
\label{fig:red_blue-Kepler}
\end{figure}

The PRM plot shown in Fig. \ref{fig:red_blue-Kepler} highlights a clear trend, where the ``red'' and ``other'' components are more asymmetric than the blueshifted parts, with the exception of the ``other'' Si component. W note that the centroid used in the PRM calculation is the centroid from the elements' total emission, which in the case of Kepler offsets it significantly to the north of the geometrical center. The ``blue'' and ``other'' emission are dominated by the northern rim, thus more offset than the ``red'' and total line emission for Si, S, Ar, and Ca. The Fe line, which shows emission closer to the core, does not follow the exact same trend.

\begin{figure}
\centering
\includegraphics[width=8.7cm]
{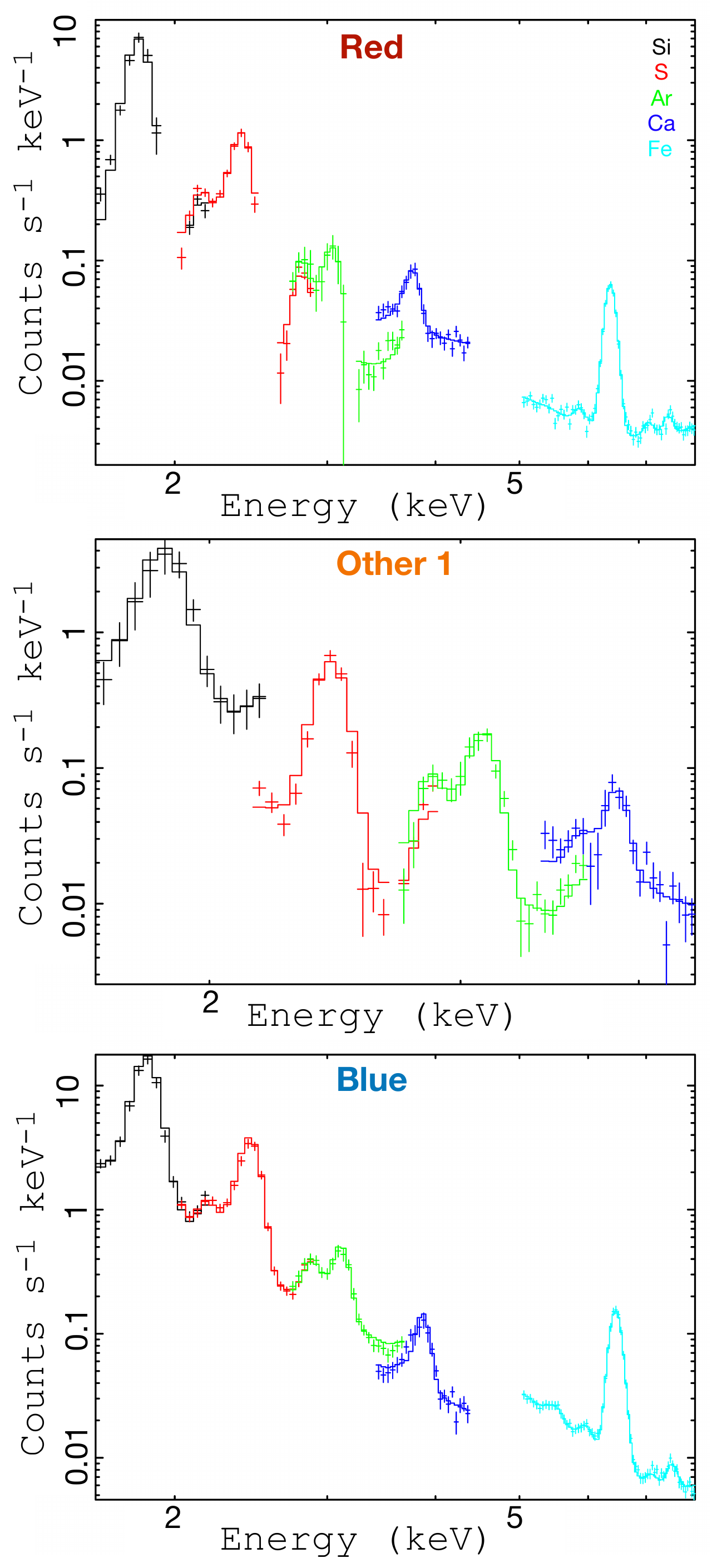}
\caption{{\it Xspec} fits of the spectra retrieved by pGMCA in Kepler. The parameter values and a description of the fittings can be found in Table \ref{tab:fits-xspec}.
}
\label{fig:Kepler-xspec}
\end{figure}

\cite{Katsuda_2015} estimated that $\sim 87\%$ of the outer ejecta is shocked, while an estimate based purely on the shocked and unshocked volumes gives a value of $\sim 60\%$ \citep{hollandashford2023estimating}. In both cases, 
the ratios from Table \ref{tab:frac-Kepler} show that most of the ejecta emission (used as a proxy for ejecta mass) is slightly blueshifted (velocity estimated: -540 $\pm$ 120 km s$^{-1}$), while a small part is significantly redshifted (velocity estimated: $\sim$6900 $\pm$ 330 km s$^{-1}$ for Si-Ca and $\sim$3000 $\pm$ 600 km s$^{-1}$ for Fe). This result is consistent with conservation of momentum; while most of the ejecta mass is moving north, a smaller portion is ejected directly opposite us, along the line of sight. \cite{Vink_2008} also found the proper motion of the northern rim to be low, which is indicative of a low tangential velocity. 

The low velocity of the northern rim might be a sign of a relatively recent interaction with CSM, which is consistent with previous studies reporting a north-south density gradient of an order of magnitude in Kepler's surrounding \citep{Blair_2007,Williams_2012}. Although an interaction with the CSM would favour a single-degenerate scenario \citep{Patnaude_2012}, a companion star is yet to be found \citep{Kerzendorf_2014,Sato_2017,Ruiz-Lapuente_2018}.


As described in Section~\ref{sect:image}, we performed spectral fitting to ensure that 
the labels we attributed to the components retrieved by pGMCA in Kepler's SNR were meaningful. 
For our ``red'' pGMCA components in Kepler's SNR, our spectral fits returned ionization timescales of a few times $10^{10}$ s cm$^{-3}$ for the Si, S, Ar, \& Ca emission, and $\sim$1 $\times 10^{10}$ for Fe ejecta. Our best-fit redshifts corresponded to velocities of 6900 $\pm$ 300 km s$^{-1}$ for the Si--Ca emission and 3000 $\pm$600 km s$^{-1}$ for the Fe-component (90\% confidence intervals; see Fig.~\ref{fig:Kepler-xspec} and Table~\ref{tab:fits-xspec}). We conclude that these pGMCA components are strongly redshifted, matching our labels. These high velocities match with our 2D maps, as a large portion of the emission is from the knot of emission in the center of the SNR: a region that should have a low tangential velocity but a high line-of-sight velocity. Our estimated Fe velocity matches well with the 2531$^{+289}_{-290}$ km s$^{-1}$ ejecta velocity of the ``CS-E Center'' region reported by \cite{kasuga21} and the $\sim$500-3200 km s$^{-1}$ for an ejecta knot near this central SE knot reported by \cite{millard20}, confirming its redshift. 

For our ``other'' pGMCA components in Kepler's SNR, our spectral fits returned an ionization timescale of 1.2 $\times 10^{11}$ s cm$^{-3}$ and a redshift corresponding to a line-of-sight velocity of 420$^{+150}_{-20}$ km s$^{-1}$: slightly redshifted (see Fig.~\ref{fig:Kepler-xspec} and Table~\ref{tab:fits-xspec}). A low velocity for this region makes sense as emission in these components is mostly coming from the northern rim of Kepler's SNR, and thus should have a high tangential but low line-of-sight velocity. Importantly, our fits prove that this pGMCA component has a different Doppler-shifted velocity than our labeled ``red'' components. \cite{millard20} found small velocities for ejecta in the northern rim, consistent with zero or slight (hundreds of km s$^{-1}$) doppler shifts.

For our ``blue'' pGMCA components in Kepler's SNR, our spectral fits returned ionization timescales of $\sim 1 \times 10^{11}$ s cm$^{-3}$ for the Si, S, Ar, \& Ca emission, and $\sim$3 $\times 10^{10}$ for Fe ejecta. Our fits produced redshifts corresponding to velocities in the range of -720 to 100 km s$^{-1}$ for the Si--Ca emission and -480 to 660 km s$^{-1}$ for the Fe-component (90\% confidence intervals; see Fig.~\ref{fig:Kepler-xspec} and Table~\ref{tab:fits-xspec}). These values are significantly lower than the Doppler shifts in our ``red''-shifted material and their centers are slightly lower than the velocities of our pGMCA components labeled ``other 1.''   Visually, there is a lot of overlap in our 2D ejecta maps for our ``other 1'' and ``blue'' pGMCA components, but the latter have more emission in the western and central regions of Kepler's SNR. Again, it makes sense that our measured Doppler-shifted velocities are small because the emission in these components is mainly in the outer rim of Kepler's SNR. \cite{kasuga21} reported edge-ejecta velocities of -300 to 700 km s$^{-1}$, matching our findings. Our findings of high redshifted velocities and only moderate blueshifted velocities match the findings of \cite{millard20}, supporting the existence of bulk asymmetry in the ejecta of Kepler's SNR .

\subsection{Tycho}
\label{sect:Tycho}

In Tycho as well, the total line emission images from Fig.~\ref{fig:red_blue_images-TOT} appear very similar to each other, with the exception of Fe that shows a more compact distribution. 
The shifted images from Fig.~\ref{fig:red_blue_images-Tycho} show also a great similitude between the differently shifted components from Si and S. For Ar and Ca, the algorithm could not retrieve as many components, so the results appear to be a mix of the different emission extracted from Si and S. In particular, the ``other 1'' and ``other 2'' components from Ar are endowed with poorly reconstructed spectra that we could not use in our spectral analysis, so they were labeled uniquely on the basis of their morphological distribution.

\begin{table}
\centering
\begin{tabular}{c|c c c c}
  \hline
  \hline
    & Redshifted  & Other 1 & Other 2 & Blueshifted \\
  \hline
  Si & 0.23 & 0.40 & 0.25 & 0.12 \\
  
  S & 0.11 & 0.29 & 0.50 & 0.10 \\
  
  Ar & - & 0.11 & 0.10 & 0.79 \\
  
  Ca & 0.22 & - & - & 0.78 \\
  
  Fe & - & 0.21 & 0.57 & 0.22 \\
  \hline
\end{tabular}
\caption{Fractions of the counts in the total image that belong to the redshifted, blueshifted, or the ``other'' parts, for each line in Tycho.}
\label{tab:frac-Tycho}
\end{table}

The most blueshifted component appears as a collection of clumps mainly present in the north, while the most extreme redshifted component appears a collection of evenly distributed clumps throughout the SNR. A detailed study of the proper motion of these ejecta clumps was carried out with GMCA in \cite{godinaud2023fresh}. In Si and S, the ``other 1'' components look like parenthesis-shaped rims surrounding the remnant east and west, while the ``other 2'' components mainly consist of a bright clump in the northwest. We only found two components in the Ca emission and three in the Ar, and their blueshifted emission appears like a mix from the ``blue,'' ``other 1,'' and ``other 2'' components from Si and S. The ``other 1'' and ``other 2'' components from Ar appear similar to those from Si and S. As the spectra appear especially poorly reconstructed for Ar, we can infer that the pGMCA struggled to properly disentangle them. The Fe components appear significantly different. Most of the blueshifted emission appears to be in a small clump in the east previously studied by \cite{Yamaguchi_2017}, while the ``other 1'' and ``other 2'' components appear as broad rims in the southwest and northwest of the remnant, respectively. There is no proper redshifted emission.

\begin{figure}
\centering
\includegraphics[width=9cm]
{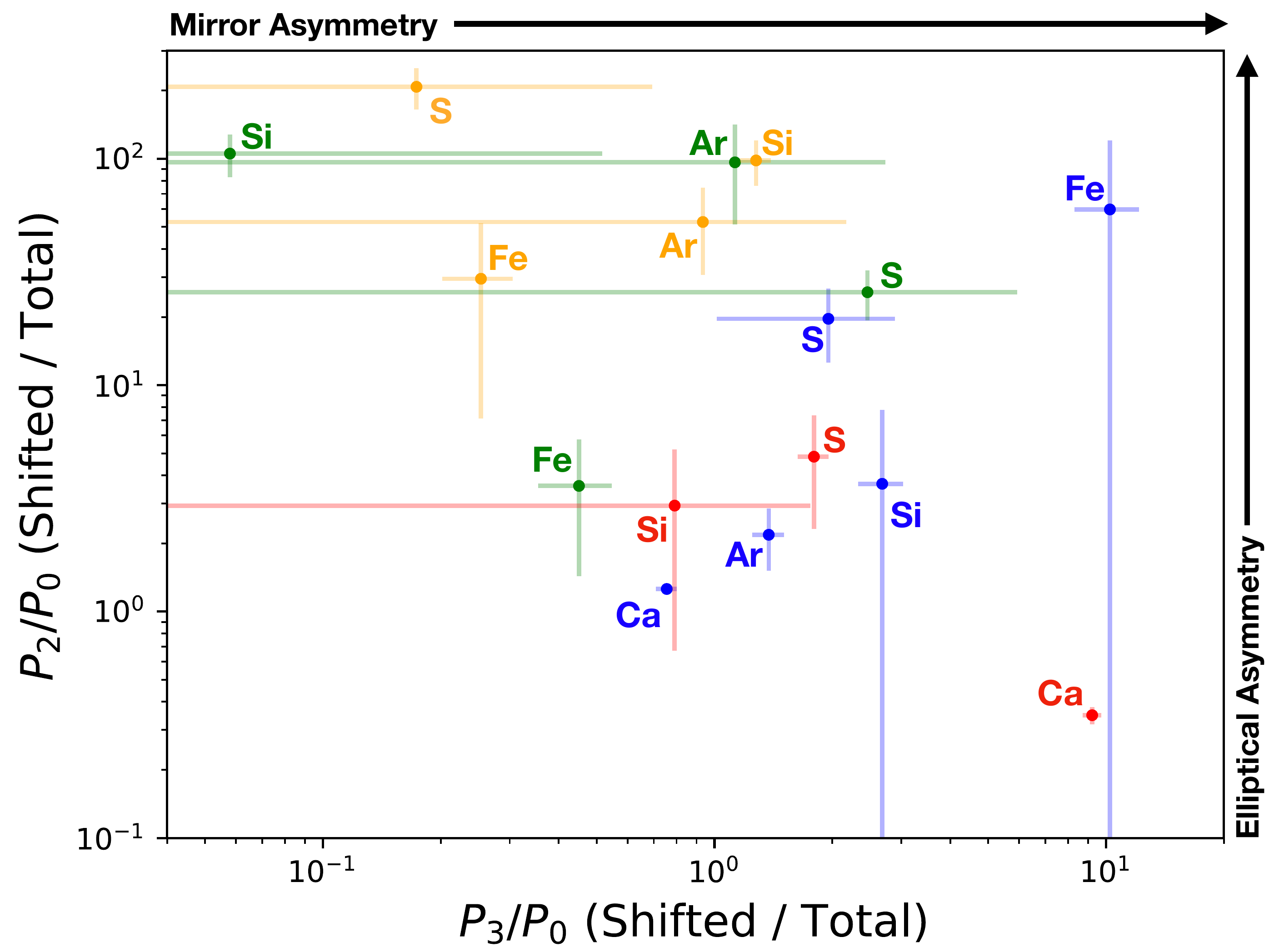}
\caption{Quadrupole power-ratios $P_{2}/P_{0}$ versus the octupole power-ratios $P_{3}/P_{0}$ of the redshifted, blueshifted, and ``other'' images of the different line emission in Tycho shown in Fig. \ref{fig:red_blue_images-Tycho}, normalized with the quadrupole and octupole power-ratios of the total images. The dots represent the values measured for the pGMCA images obtained from the real data, and the error bars the standard deviation of fifty selected outputs from pGMCA on constrained bootstrap resamplings.}
\label{fig:red_blue-Tycho}
\end{figure}

The PRM plot shown in Fig. \ref{fig:red_blue-Tycho} does not highlight any trend. Table \ref{tab:frac-Tycho} shows that most of the emission is in the intermediate components ``other 1'' and ``other 2'' for Si, S, and Fe. For Ar and Ca, most of the emission is in the blueshifted component, that appears like a mix of the ``blue'', ``other 1,'' and ``other 2'' emission from Si and S. The ``other 2'' component is largely dominated by a small clump in the northwest, but other components show a roughly spherical distribution.

\begin{figure*}
\centering
\includegraphics[width=17cm]
{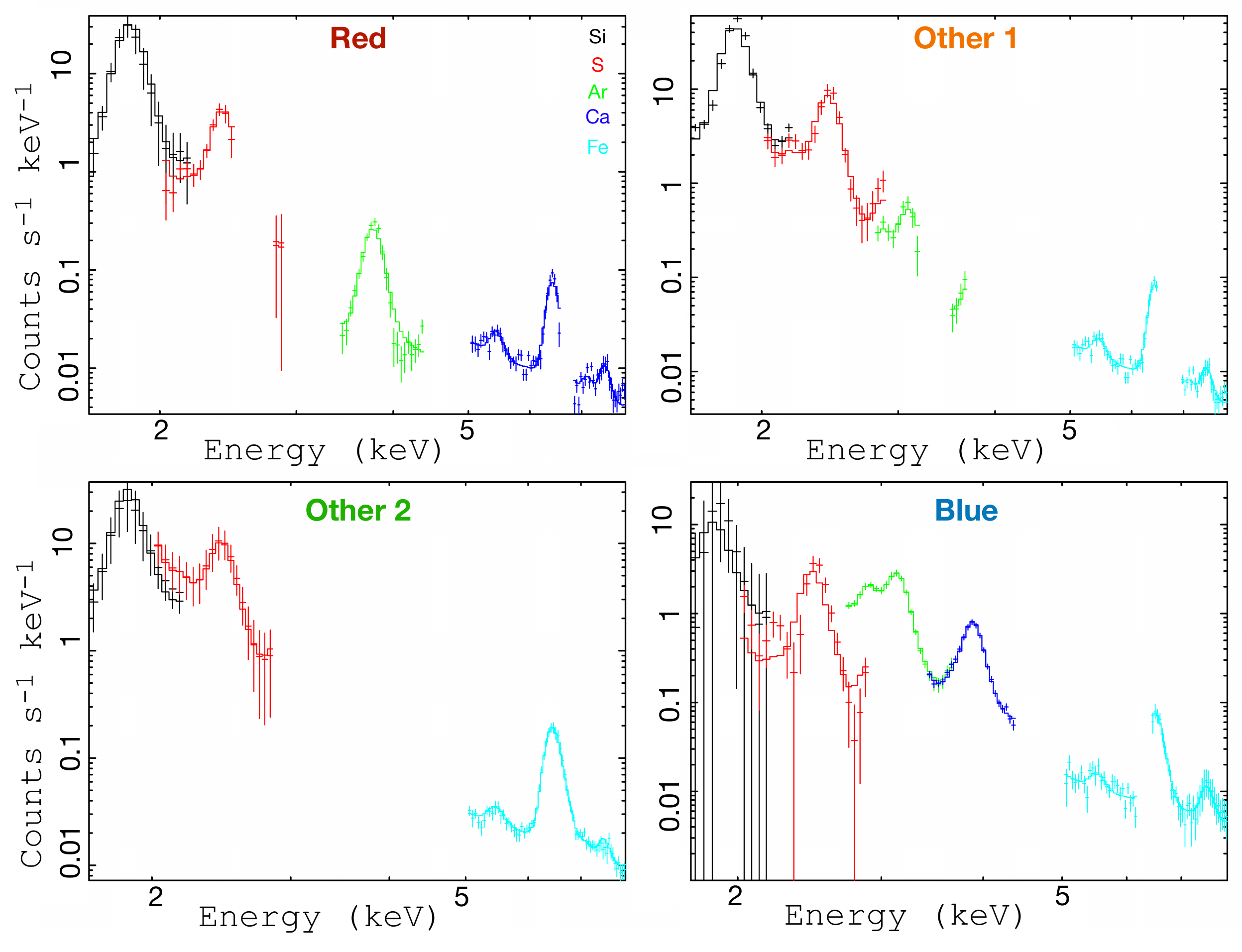}
\caption{{\it Xspec} fits of the spectra retrieved by pGMCA in Tycho. The parameter values and a description of the fittings can be found in Table \ref{tab:fits-xspec}.
}
\label{fig:Tycho-xspec}
\end{figure*}

\cite{williams17} and \cite{sato17} performed ejecta velocity measurements of Tycho's SNR using {\it Chandra} {\sc ACIS} observations. Our distribution maps seem to roughly match their findings: the blueshifted ejecta are mainly present in the center-east and north, while the redshifted ejecta are present in the center-west and south; we both find ionization timescales to be in the range of $\sim10^{9}-10^{11}$ s cm$^{-3}$ and $\sim$1.0--few~keV electron temperatures. \cite{sato17} found velocities of around -3000 to -4000 km s$^{-1}$ for the blueshifted Si ejecta, and velocities of -1000 to -2500 km s$^{-1}$ km s$^{-1}$ for the eastern knot. Our spectral fits report similarly strong blueshfited velocities: -1600 $\pm$ 100 km s$^{-1}$ for the Si \& S emission and -2400 $\pm$ 1100 km s$^{-1}$ for the Ar, Ca, and Fe emission: emission dominated by the eastern knot (90\% CI; see Fig.~\ref{fig:Tycho-xspec} and Table~\ref{tab:fits-xspec}). As our maps have most of their emission on the edges of Tycho's SNs -- namely, regions with lower line-of-sight  velocities, but
higher tangential ones --- it makes sense that our measured Doppler shifts are smaller in magnitude. Thus, it is reasonable to conclude that our ``blue'' labeled pGMCA components are accurately capturing ejecta moving towards us. These results are also consistent with the line-of-sight velocities found by \cite{godinaud2023fresh}, which were retrieved in a similar way using GMCA components. 

The ``low-velocity ejecta blobs'' from \cite{sato17} are around the edges of Tycho's SNR and should thus be captured by our ``other1'' components. Our fits to Si, S, and Ar emission produce best-fit 
Doppler velocities of 780--1530 km s$^{-1}$ and our fits to Fe produce Doppler velocities of -1530 to 210 km s$^{-1}$ (90\% CI; see Fig.~\ref{fig:Tycho-xspec} and Table~\ref{tab:fits-xspec}). 
These velocities are moderate to lower in magnitude than those in our ``blueshifted'' ejecta maps. Interestingly, even though our ``Other 1'' Fe pGMCA component displays emission clearly to the left of the Fe K-$\alpha$ complex, we find that it is slightly blueshifted. We note that the pGMCA FeK-$\alpha$ spectral line complex is visially cutoff when plotted, showing a sharp drop-off and small line width that does not match up with any plasma model we tried. It is possible that this emission is part of a larger plasma region (perhaps, in combination with the similarly cutoff ``Fe-Blue'' component) that is all blueshifted.

Our ``other 2'' components are dominated by a northwestern knot and more diffuse emission. Our Si \& S pGMCA components are best fit by a plasma with a moderate ($\sim$5.5e10 cm s$^{-3}$) ionization timescale and our Fe pGMCA component is best fit by a plasma with a low (few times 10$^9$ s$^{-3}$ cm s$^{-3}$) ionization timescale. However, the best-fit Doppler shifts were very uncertain, settling on a spread of values from different fits. The Si \& S emission had Doppler shifts corresponding to -3600 to 2100 km s$^{-1}$, while the Fe emission had a Doppler shift corresponding to -600 to 2400 km s$^{-1}$ (90\% CI; see Fig.~\ref{fig:Tycho-xspec} and Table~\ref{tab:fits-xspec}). These are consistent with zero or moderate blue- and redshifts, possibly greater in magnitude than our ``other 1'' emission. The large spread and consistency with low-to-moderate Doppler shifts is consistent with our 2D maps; there seems to be more emission from both the center and rim of Tycho's SNR in these components, representing the NW $\sim$half of Tycho's SNR in Si \& S emission as opposed to only the rim.



For Si--Ca, ``red'' labeled components (excluding Ar as its spectra are too poor in quality), we find extreme velocities of 4500--9500 km s$^{-1}$ (90\% CI; see Fig.~\ref{fig:Tycho-xspec} and Table~\ref{tab:fits-xspec}). These velocities closely match the 4000--7500 km s$^{-1}$ velocities found by \cite{sato17}, providing support that our ``red'' labeled pGMCA components are indeed capturing ejecta moving away from us.

Additionally, our pGMCA components for Tycho's SNR generally match up with a previous GMCA analysis on Si emission in Tycho's SNR \citep{godinaud2023fresh}. They found that the redshifted emission is largely present in the southern hemisphere of the SNR, with central velocities for both red- and blueshifted ejecta of a few thousand km s$^{-1}$, as we do. Visually, our combined "ed'' and ``other 1'' pGMCA Si components look like their identified redshifted map, and our combined ``other 2'' and ``blue'' pGMCA Si components look like their identified blueshifted map.

Overall, the mostly symmetrical line-of-sight velocity distribution, the similar ejecta morphologies, and the strong quadrupole power-ratios suggest a bipolar explosion with a north and south elongation tilted towards the observer and even amounts of ejecta accelerated in each direction.

\begin{table*}
\centering
\begin{tabular}{c|c c c c}

  Parameter  & Redshifted & Other 1 & Other 2 & Blueshifted \\
  \hline
    &  & Kepler &  & 
  \\
  \hline
  \\
  $\tau_{Si-Ca}$ (s cm$^{-3}$)& 5.3e10$^{+4.3e10}_{-1.3e10}$ & 1.2e11$^{+7.4e10}_{-3.8e10}$ & - & 1.7e11$^{8.5e9}_{-1.7e10}$ \\
  \\
  $\tau_{Fe}$ (s cm$^{-3}$) & 1.1e10$^{+1.9e10}_{-4.7e-9}$ & - & - & 2.3e10$^{+3.4e9}_{-2.6e9}$ \\
  \\
  redshift$_{Si-Ca}$ & 2.3e-2$^{+1.4e-3}_{-6.3e-4}$ & 1.4e-3$^{5e-4}_{-7e-5}$ & - & 2.2e-4$^{+1e-4}_{-2.2e-3}$ \\
  \\
  redshift$_{Fe}$ & 9.9e-3$^{+1.9e-3}_{-2.5e-3}$ & - & - & -1.6e-3 to 2.2e-3 \\
  \\
  V$_{Si-Ca}$ (km s$^{-1}$) & 6900$^{+420}_{-190}$ & 420$^{+150}_{-20}$ & - & 66$^{+30}_{-660}$ \\
  \\
  V$_{Fe}$ (km s$^{-1}$) & 3000$^{+570}_{750}$ & - & - & -480 to 660 \\
  \\
  \hline
    &  & Tycho &  & 
  \\
  \hline
  \\
  $\tau_{Si-Ca}$ (s cm$^{-3}$)& 8e10 $\pm$ 3e10 & 4.3e10$^{+8.8e9}_{-5.5e9}$ & 5e10 to 7e10 & 1.07e11$^{+1.2e10}_{-1.0e10}$ \\
  \\
  $\tau_{Fe}$ (s cm$^{-3}$) & - & 6.3e9$^{+6.3e9}_{-3e9}$ & 6e8 to 7e9 & 1.77e10$^{+5.1e9}_{-5.3e9}$ \\
  \\
  redshift$_{Si-Ca}$ & 2.9e-2 $^{+2.8e-3}_{-1.2e-2}$ & 2.8e-3$^{+2.3e-3}_{-2.0e-4}$ & -1.4e-2 to 7e-3 & -5.2e-3$^{+2.2e-4}_{-3.5e-4}$ \\
  \\
  redshift$_{Fe}$ & - & -2.4e-3$^{+2.7e-3}_{-3.1e-3}$ & -2e-3 to 8e-3 & -8.2e-3$^{+3.9e-3}_{-3.4e-3}$ \\
  \\
  V$_{Si-Ca}$ (km s$^{-1}$) & 8700$^{+840}_{-4200}$ & 840$^{+690}_{-60}$  & -3600 to 2100 & -1560$^{+66}_{-105}$  \\
  \\
  V$_{Fe}$ (km s$^{-1}$) & - & -720$^{+930}_{-810}$ & -600 to 2400 & -2460$^{+1170}_{-1020}$ \\
\end{tabular}
\caption{Results of fitting the spectral components retrieved by pGMCA. In Kepler, we see that the measured Doppler shifts are all consistent with positive (redshifted) velocities. Our ``red'' component is highly redshifted, while the other components have small velocities. For the Fe velocity in the ``blue'' component, multiple equally-valid fits returned different best-fit Doppler shifts; we report the spread between these models.
\\
In Tycho, we see that our measured Doppler shifts span large positive and negative velocities. Our ``red'' component is highly redshifted, our ``blue'' component is moderately blueshifted, and our ``other 1'' component shows a mix of both slight redshifts and blueshifts. As noted in Section~\ref{sect:Tycho}, the apprent cut-off of the Fe K-$\alpha$ line in this component suggest that this ``other 1'' Fe emission might actually be part of the ``blue'' component. As for the ``other 2'' Tycho component, multiple equally valid fits returned different best-fit parameters; we report the spread between these models instead of the 90\% CI from a single model.}
\label{tab:fits-xspec}
\end{table*}

\section{Comparison with Cassiopeia A}
\label{sub:comparison}

\begin{figure}
\centering
\includegraphics[width=9cm]
{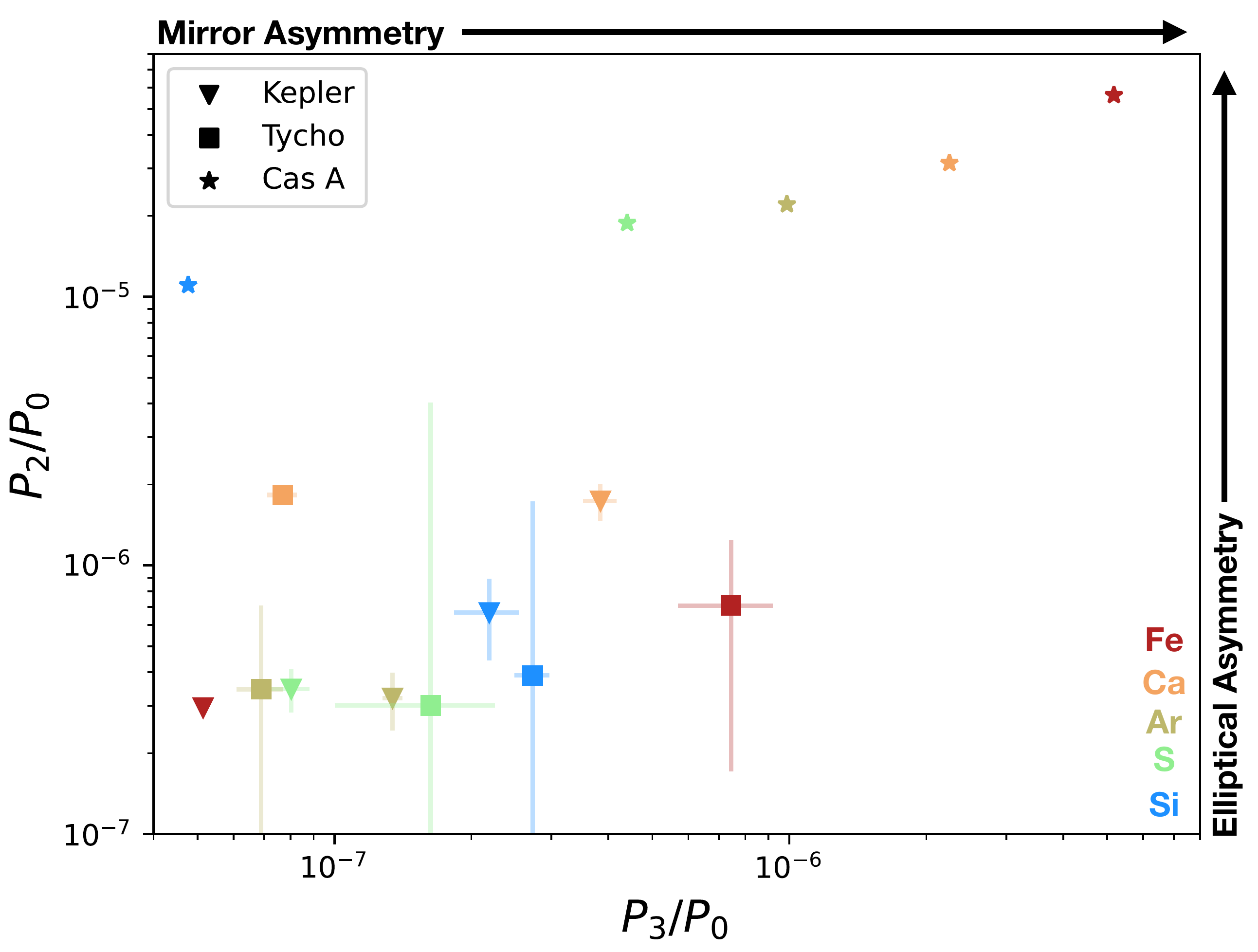}
\caption{Quadrupole power ratios, $P_{2}/P_{0}$, versus the octupole power-ratios, $P_{3}/P_{0}$, of the total images of the different line emission in Kepler and Tycho. The dots represent the values measured for the pGMCA images obtained from the real data, and the error bars the standard deviation of fifty selected outputs from pGMCA on constrained bootstrap resamplings.} 
\label{fig:asym-Kepler-Tycho}
\end{figure}

The PRM plot in Fig. \ref{fig:asym-Kepler-Tycho} shows a comparison of the asymmetries of the total line emission in Kepler and Tycho, to which we added the PRM values we found for Cassiopeia A in \cite{Picquenot_2021}. It appears that the total line emission in Cassiopeia A are all significantly more elliptical asymmetric than those in Kepler and Tycho and mostly more mirror-asymmetric. This is consistent with the results of \cite{lopez09b}, highlighting that core collapse SNRs had an overall morphology that tended to be more asymmetric than Type Ia SNRs.

In the case of Cassiopeia A, we had noted in our previous paper that heavier elements were ejected more asymmetrically than lighter elements, which is consistent with core-collapse simulations \citep{wongwathanarat13,janka17,Gessner_2018,M_ller_2019}. Here, we can notice that Kepler and Tycho do not follow the same trend. Furthermore, Si, S, and Ar even follow a reverse trend in both Tycho and Kepler, were the lighter elements are the most asymmetric. The Fe in Kepler appears to be the most symmetrically distributed element and the Ca in Tycho is one of the two most mirror-symmetric elements. Interestingly, the Ca emission is the most elliptically asymmetric element in both remnants. With a sample size in this work of only two objects, it is premature to draw specific conclusions regarding the link between asymmetries and elemental atomic numbers in Type Ia SNRs. Subsequent studies, with the application of the same method on a larger number of SNRs would allow us to explore this question more thoroughly.

However, simulation results predict that Fe-group elements are located in the inner parts of the ejecta, surrounded by an outer shell of intermediate-mass elements \citep[e.g., Si, S, Ca, and Ar;][]{Seitenzahl2013,Stone_2021}. Our results are consistent with this, since  the Fe emission appears more compact for both remnants and closer to the core than the other emission, while also presenting a significantly different morphology. The line-of-sight velocities are harder to compare, as our fits are endowed with large error bars. However, the velocity we retrieved are mostly compatible with larger velocities for the intermediate-mass elements than for Fe, as expected. The main discrepancy is the blueshifted component in Tycho, where $V_{Fe}$=-2460 $\pm$ 1100 km~s$^{-1}$ and we $V_{Si-Ca}$=~-1500~$\pm$~100~km~s$^{-1}$. The blueshifted Fe component mainly consists of a southeastern knot, that \cite{millard20} found to move radially at about 5000 km $s^{-1}$. This knot was thoroughly studied by \cite{Yamaguchi_2017}. They found its ionization timescale to be highly different from that of its Si-rich surrounding, favoring the hypothesis that during the explosion, this knot detached from the rest of the Fe-rich material and formed a protrusion at the edge of the ejecta.

\section{Conclusions}

By using a new methodology and applying it to the Kepler and Tycho {\it Chandra} X-ray data, we were able to revisit the mapping of the intermediate-mass elements and Fe and separate them into differently red- or blue-shifted parts, allowing us to investigate the three-dimensional (3D) morphology of these SNRs. These new maps and the associated spectra could then be used to quantify the asymmetries of each component, their mean direction and their line-of-sight velocity. The main findings of the paper are consistent with the general results found in the previous studies cited throughout our study.

In Kepler, the most blueshifted component appears as a broad rim following the forward shock in the north, generally considered  in previous studies to be trace of a CSM interaction. This component appears to have a low velocity, that makes it compatible both with a slight blueshift or with a slight redshift.  A smaller portion of the the ejecta is significantly redshifted, which is consistent with the conservation of momentum: most of the ejecta is moving slowly to the north, while a small portion is ejected directly opposite us at higher speeds. An extremely off-center explosion site \citep{Maeda10} could result in this ejecta distribution.

In Tycho, the redshifted emission appears as a collection of clumps slightly more present in the center-east and south, while the blueshifted emission seems mainly present in the north and west of the remnant. The PRM plot does not show any clear trend, and the mostly symmetrical line-of-sight velocity distribution from past studies, consistent with our own findings, suggests  a bipolar explosion with a north-south elongation tilted towards the observer, and even amounts of ejecta accelerated in each direction. Such a bipolar explosion might be the result of a double-degenerate origin due where close interactions between the white dwarfs and/or instabilities in material accreted along the orbital plane led to an asymmetric explosion.

In both remnants, the Fe emission appears more compact, closer to the core than the other emission and presents a significantly different morphology. The line-of-sight velocities we derive are endowed with large error bars, but they are mostly compatible with lower values for the Fe emission, with the notable exception of a southeastern clump in Tycho. 

A comparison with the results obtained in a previous paper on the core-collapse SNR Casiopeia A also proved meaningful. The total line emission in Cassiopeia A are all significantly more elliptical asymmetric than those in Kepler and Tycho, and mostly more mirror asymmetric, which is consistent with past studies stating that the global morphology of core collapse SNRs tended to be more asymmetric than that of Type Ia. 

In Cassiopeia A, heavier elements are ejected more asymmetrically than lighter elements, which is consistent with core-collapse simulations. Kepler and Tycho's SNRs, however, seem to exhibit almost a reverse trend, where many heavier elements present more symmetric morphologies than lighter elements in these SNRs. This might be a result of the detonation processes and/or pre-explosion dynamics in Type Ia SNe; for instance, an initial detonation of an outer He-shell in the ``double-detonation'' mechanism or a differing number of explosion kernels (potentially off-center) for the central explosion (e.g., \citealt{Maeda10,seitenzahl13}) could result in these observables.
We note that a sample of two SNRs is too small to conclude anything and suggest that studies, with the application of the same method on a larger number of targets, would enable exploration of this question more thoroughly. More than ten SNRs already benefit from exposure above 250 ks from \textit{Chandra} and several millions of photons detected, making them great candidates for  pGMCA applications. Good targets with deep observations would include: the three large Magellanic cloud young SNRs (0509-67.5, 0519-69.0, and N103B, each $\sim$400 ks from 2017-2020 observations) and SN 1006 ($\sim$500 ks) for Type Ia; G292.0+1.8 ($\sim$1 Ms), N132D ($\sim$1 Ms), and W49B ($\sim$270 ks total) for core-collapse.

\begin{acknowledgements}
The material is based upon work supported by NASA under award number 80GSFC21M0002. 
\end{acknowledgements}

%
   \bibliographystyle{aa} 
   \bibliography{asymmetries_kepler-tycho} 
%

\end{document}